\documentclass[journal]{IEEEtran}

\usepackage{caption}
 \usepackage{subcaption}
\ifCLASSINFOpdf
  \usepackage[pdftex]{graphicx}
  \graphicspath{{./graph/}}
  \DeclareGraphicsExtensions{.pdf,.jpeg,.png}
\else
  \usepackage[dvips]{graphicx}
  \graphicspath{{./graph/}}
  \DeclareGraphicsExtensions{.eps}
\fi

\ifCLASSINFOpdf
\else
\fi

\usepackage[cmex10]{amsmath,mathtools}
\usepackage{mdwmath}
\usepackage{mdwtab}
\usepackage{amssymb}
\usepackage{tabularx}
\usepackage{float}
\usepackage{multirow}
\usepackage{color}
\usepackage{multicol}

\usepackage{algorithmic}
\usepackage{algorithm}
\usepackage{float}

\allowdisplaybreaks[4]
\makeatother
\usepackage{epstopdf}
\usepackage{color}

\usepackage[utf8]{inputenc}

\usepackage{cite}

\begin{document}

\title{Towards Resilient Federated Learning in CyberEdge Networks: Recent Advances and Future Trends}

\author{
Kai~Li, \textit{Senior Member, IEEE,}
Zhengyang Zhang, 
Azadeh Pourkabirian, 
Wei~Ni, \textit{Fellow, IEEE,}
Falko~Dressler, \textit{Fellow, IEEE,}
and 
Ozgur~B.~Akan, \textit{Fellow, IEEE}

\thanks{K.~Li is with the School of Electrical Engineering and Computer Science, TU Berlin, Germany, and also with Real-Time and Embedded Computing Systems Research Centre (CISTER), Porto 4249--015, Portugal (E-mail: kaili@ieee.org).}
\thanks{Z.~Zhang is with the Division of Electrical Engineering, Department of Engineering, University of Cambridge, CB3 0FA Cambridge, U.K. (E-mail: zz420@cam.ac.uk).}
\thanks{A.~Pourkabirian is with Real-Time and Embedded Computing Systems Research Centre (CISTER), Porto 4249--015, Portugal (E-mail: azadeh.pourkabirian@cister-labs.pt).}
\thanks{W.~Ni is with the Digital Productivity and Services Flagship, Commonwealth Scientific and Industrial Research Organization (CSIRO), Sydney, NSW 2122, Australia (E-mail: wei.ni@ieee.org).}
\thanks{F.~Dressler is with the School of Electrical Engineering and Computer Science, TU Berlin, Germany (E-mail: dressler@ccs-labs.org).}
\thanks{O. B. Akan is with the Division of Electrical Engineering, Department of Engineering, University of Cambridge, CB3 0FA Cambridge, U.K., and also with the Center for NeXt-Generation Communications (CXC), Ko\c c University, 34450 Istanbul, Turkey (E-mail: oba21@cam.ac.uk).}
}

\IEEEcompsoctitleabstractindextext{%
\begin{abstract}
\boldmath In this survey, we investigate the most recent techniques of resilient federated learning (ResFL) in CyberEdge networks, focusing on joint training with agglomerative deduction and feature-oriented security mechanisms. We explore adaptive hierarchical learning strategies to tackle non-IID data challenges, improving scalability and reducing communication overhead. Fault tolerance techniques and agglomerative deduction mechanisms are studied to detect unreliable devices, refine model updates, and enhance convergence stability. Unlike existing FL security research, we comprehensively analyze feature-oriented threats, such as poisoning, inference, and reconstruction attacks that exploit model features. Moreover, we examine resilient aggregation techniques, anomaly detection, and cryptographic defenses, including differential privacy and secure multi-party computation, to strengthen FL security. In addition, we discuss the integration of 6G, large language models (LLMs), and interoperable learning frameworks to enhance privacy-preserving and decentralized cross-domain training. These advancements offer ultra-low latency, artificial intelligence (AI)-driven network management, and improved resilience against adversarial attacks, fostering the deployment of secure ResFL in CyberEdge networks.
\end{abstract}
}

\maketitle

\IEEEdisplaynotcompsoctitleabstractindextext
\IEEEpeerreviewmaketitle

\section{Introduction of Federated Learning in CyberEdge Networks}

\subsection{Background}
CyberEdge networks provide an advanced networking paradigm that integrates Mobile Edge Computing (MEC) and Internet of Things (IoT) technologies to provide seamless, low-latency, and high-bandwidth connectivity for users in immersive digital environments, such as the Metaverse~\cite{li2022internet}. By leveraging edge computing, CyberEdge networks process and store data closer to the user, reducing latency and improving real-time interactions for applications in Augmented Reality (AR), Virtual Reality (VR), and Mixed Reality (MR)~\cite{ergenc2025resilience}. The integration of IoT enables dynamic data exchange between users' devices, wearables, and sensors, further enhancing contextual awareness and adaptive resource management. This architecture ensures a responsive and scalable network infrastructure, supporting the high computational and communication demands of next-generation connected experiences.

Protecting data privacy while addressing bandwidth limitations is critical in CyberEdge networks, as IoT devices facilitate real-time, immersive experiences with AR/VR applications in the Metaverse~\cite{qiao2024resources,yu2023socially,xu2023learning}. These environments generate vast amounts of sensitive personal data, including biometric information, location, and interaction patterns, making them prime targets for cyber threats and unauthorized access~\cite{li2025ovp,pan2024privacy,yin2021privacy}. Traditional cloud-based data processing models require large-scale data transmission, which not only increases privacy risks but also strains network bandwidth, resulting in latency issues that degrade user experience. 

\begin{figure}[h]
\centering
\includegraphics[width=\linewidth]{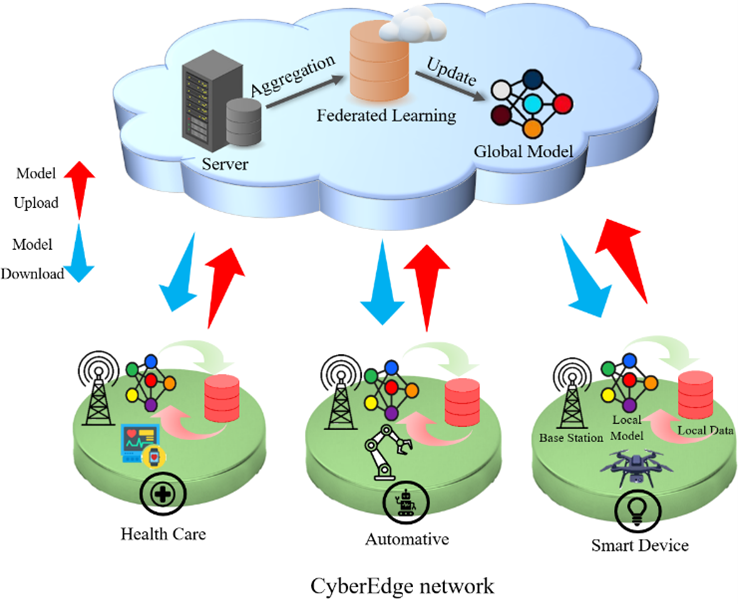}
\caption{FL in CyberEdge networks: The aggregation process}
\label{figure_application}
\end{figure}

To protect data privacy while addressing bandwidth limitations, federated learning (FL) is widely adopted in CyberEdge networks as a privacy-preserving and bandwidth-efficient machine learning paradigm~\cite{chen2024federated}. As depicted in Fig.~\ref{figure_application}, instead of uploading raw data to servers, FL enables local model training on user devices, with only model updates (e.g., weight adjustments) being shared with the server or aggregated at the edge~\cite{li2025novel,sun2024fedcpd,pei2022tkagfl}. This approach protects user privacy by keeping personal data on local devices while reducing bandwidth consumption, as significantly less data is transmitted compared to machine learning in traditional IoT systems. By enabling distributed intelligence without centralized data collection, FL enhances the scalability, responsiveness, and security of CyberEdge networks~\cite{deng2024communication,wang2023mitigating}, making them more efficient for real-time, connected applications in the Metaverse.


With the rise of emerging cyber threats, FL faces a critical resilience challenge, including poisoning attacks, adversarial manipulations, model inversion, and Byzantine attacks~\cite{wen2023byzantine,olowononi2021federated,chen2024multicenter}. Since FL relies on distributed devices for training and model updates, malicious participants can inject tampered data or compromised model updates, degrading model performance and leading to biased or incorrect predictions~\cite{ads2024rare}. Moreover, FL is susceptible to communication failures, resource constraints, and data heterogeneity, further affecting its reliability in real-world deployments~\cite{gufran2023fedhil}. In CyberEdge networks, where FL is expected to support real-time~\cite{zhu2022resilient}, privacy-sensitive applications in AR/VR, and the Metaverse~\cite{wan2024data}, ensuring its resilience is paramount. A compromised FL system not only threatens user privacy and security but also undermines the effectiveness of intelligent services that depend on it~\cite{shao2022dres}. Strengthening FL’s resilience through robust aggregation mechanisms, anomaly detection, secure communication, and trust-aware learning frameworks is crucial for maintaining the integrity, availability, and security of CyberEdge networks~\cite{ergenc2025resilience,khraisat2024survey}.

\subsection{Our Motivation}
In this paper, we investigate two emerging key techniques towards resilient FL (ResFL) in CyberEdge networks, i.e., joint training and agglomerative deduction, as well as feature-oriented threats and defenses. Specifically, joint training and agglomerative deduction can enhance FL resilience by leveraging heterogeneous data and hierarchical aggregation with fault tolerance and anomaly detection. Heterogeneous data, a fundamental challenge in FL, stems from diverse edge devices with non-independent and identically distributed (non-IID)  data distributions, requiring adaptive aggregation strategies, such as hierarchical learning, where local models are first aggregated at intermediate levels before contributing to the global model. This approach improves scalability and reduces communication overhead. Meanwhile, fault tolerance mechanisms ensure system robustness by detecting and mitigating the impact of unreliable or compromised devices through anomaly detection techniques, such as statistical analysis or machine learning-based outlier detection. An agglomerative deduction can further improve resilience by iteratively refining updates, filtering out low-quality contributions, and prioritizing reliable data sources, leading to more stable model convergence.

On the other hand, feature-oriented threats and defenses focus on securing FL models against adversarial manipulations, particularly poisoning attacks as well as inference and reconstruction attacks. In poisoning attacks, adversaries inject malicious data or manipulate local model updates to degrade global model performance. Defenses against such threats include robust aggregation techniques (e.g., Krum, median-based aggregation) and anomaly detection mechanisms that identify suspicious updates based on model divergence. Inference and reconstruction attacks exploit model updates to infer sensitive training data or reconstruct private features, threatening data confidentiality. Defense strategies, such as differential privacy, homomorphic encryption, and secure multi-party computation, mitigate these risks by obfuscating updates and limiting information leakage. Addressing both poisoning and inference threats, feature-oriented security mechanisms can enhance the robustness and privacy of ResFL in CyberEdge networks.

Furthermore, we investigate several key opportunities and future research directions for constructing ResFL in CyberEdge networks. The evolution of 6G brings ultra-low latency, massive connectivity, and AI-native infrastructure that can significantly accelerate the deployment of ResFL. Research can focus on optimizing FL for 6G by leveraging intelligent resource allocation, semantic communication, and dynamic edge-cloud collaboration. Network-aware FL and federated reinforcement learning can be developed to support self-optimizing systems capable of real-time adaptation to network conditions, mobility, and security demands, enhancing overall resilience and performance.

Another promising direction lies in integrating Large Language Models (LLMs) and enabling collaborative cross-domain and cross-silo ResFL. LLMs introduce opportunities for privacy-preserving and decentralized training on edge devices, especially when enhanced with model compression, personalized FL, and secure protocols that guard against adversarial threats. At the same time, cross-domain collaboration supported by interoperable learning frameworks allows diverse sectors, such as healthcare, transportation, and smart infrastructure, to contribute to and benefit from shared intelligence while preserving data privacy and regulatory compliance. These future directions and insights pave the way for building robust, scalable, and trustworthy ResFL systems that can adapt to heterogeneous environments and empower next-generation CyberEdge applications.

\subsection{Contributions}
The key contributions of this paper are as follows: 
\begin{itemize}
    \item We study the joint training and agglomerative deduction techniques in CyberEdge networks, which aim to improve ResFL by leveraging heterogeneous data, hierarchical aggregation, fault tolerance, and anomaly detection. To address non-IID data challenges, we explore adaptive hierarchical learning strategies that improve scalability and reduce communication overhead. We also present fault tolerance and agglomerative deduction mechanisms that detect unreliable devices, refine model updates, and prioritize high-quality contributions for stable convergence.
    \item Unlike existing literature on FL security, we investigate the new feature-oriented threats and defenses in ResFL, focusing on poisoning attacks, inference attacks, and reconstruction attacks that utilize benign model features to compromise model integrity and data privacy. To counter poisoning threats, we explore resilient aggregation techniques and anomaly detection for identifying and filtering malicious updates, such as differential privacy, homomorphic encryption, and secure multi-party computation, which can enhance the ResFL in CyberEdge networks.
    \item We explore opportunities and future research directions for constructing ResFL in CyberEdge networks, emphasizing the integration of 6G and LLMs. The advancement of 6G brings ultra-low latency, massive connectivity, and AI-driven network management, while LLMs enable privacy-preserving and decentralized training across edge devices. These innovations are crucial for accelerating ResFL deployment and enhancing security against data leakage, adversarial manipulation, and backdoor attacks.
\end{itemize}

\subsection{Paper Structure}
The rest of this survey is organized as follows. In Section~\ref{sec_ii}, we examine the gaps in existing surveys about reliable and secure FL. 
Section~\ref{sec_iii} studies joint training and agglomerative deduction technologies that enhance FL resilience by leveraging heterogeneous data and hierarchical aggregation with fault tolerance and anomaly detection. 
Section~\ref{sec_iv} studies feature-oriented threats and defenses that focus on securing FL models against adversarial manipulations, particularly poisoning attacks as well as inference and reconstruction attacks. 
The research opportunities for building future ResFL in CyberEdge networks are delineated in Section~\ref{sec_v}. 
Section~\ref{sec_vi} concludes the survey.

\section{Related Work}
\label{sec_ii}
In this section, we review the literature thoroughly in terms of the reliability and security of FL in CyberEdge networks.

\subsection{Reliable Federated Learning}
Recent advancements in FL have led to diverse approaches to enhance the reliability and robustness of distributed systems, particularly when integrated with edge computing and IoT applications. For instance, a federated edge architecture incorporating semantic IoT was studied by Li et al.~\cite{li2023towards}, enabling AR/VR users to offload resource-intensive semantic processing tasks to edge servers. These servers collaboratively train a unified semantic model using FL-based frameworks. To ensure reliability and efficiency, the researchers developed a dynamic sequential-to-parallel FL approach, incorporating semantic compression and compensation techniques. This strategy can merge compressed historical semantic data and fine-tune classifier parameters, thus optimizing resource usage and model accuracy.

Addressing FL reliability from a security perspective, Murmu et al.~\cite{murmu2024reliable} introduced a customized, inequality-aware FL designed specifically for secure color image transmission within CyberEdge networks. Their personalized approach adapts data sampling algorithms to client-specific requirements based on the local availability of labeled data. Complementing these personalized FL efforts, Kang et al.~\cite{kang2020reliable} studied a reputation-based metric to select trusted workers. By leveraging participants' reputation values, their framework filters unreliable clients, enhancing the accuracy and trustworthiness of the learning process.

Several surveys have broadened the understanding of reliable FL and identified critical challenges across various sectors. Nguyen et al.~\cite{nguyen2022federated} presented an extensive survey emphasizing reliable FL's applications in smart healthcare, including federated electronic health records management, remote health monitoring, medical imaging, and COVID-19 detection. Their work outlined motivations, technical prerequisites, and opportunities for further deployment in healthcare systems. In addition, Huang et al.~\cite{huang2024federated} provided a review of FL techniques focused on three aspects: generalization, robustness, and fairness. They categorized existing methods based on distinct task settings, such as cross-client and out-client shifts in generalizable FL, Byzantine attacks, reward conflicts, and prediction biases. Their work also highlighted data heterogeneity as an ongoing critical challenge that demands targeted future research.

Meanwhile, Gabrielli et al.~\cite{gabrielli2023survey} and Jiang et al.~\cite{jiang2024blockchained} offered another perspective by categorizing reliable FL schemes into two primary groups: traditional distributed computing-based FL and blockchain-integrated FL. Their analyses identified significant challenges, including vulnerabilities to adversarial attacks and the absence of effective incentive mechanisms to encourage participation.

Complementing these insights, Khan et al.~\cite{khan2021federated} evaluated FL specifically tailored for IoT applications, focusing on crucial metrics like scalability, quantization, and security. Their survey provided a taxonomy addressing system parameters, federated optimization schemes, incentive mechanisms, security measures, and operational modes. Building upon this, Boobalan et al.~\cite{boobalan2022fusion} reviewed the integration of FL with industrial IoT, discussing critical aspects such as privacy preservation, resource management, and efficient data handling. They discussed the motivations and benefits of combining FL with industrial IoT, emphasizing privacy protection and enabling on-device learning capabilities.

Furthermore, the structural considerations and impacts of network topologies on FL effectiveness were explored by Wu et al.~\cite{wu2024topology}, who revealed that certain network topologies introduce additional constraints and opportunities in FL systems. For example, employing ring topology can significantly improve scalability and accommodate diverse client activities, all while eliminating dependency on a central server.

\subsection{Secure Federated Learning}
FL continues to gain prominence due to its inherent potential to address vulnerabilities stemming from increased interconnectivity, data exchange, and digital transformation. In the survey~\cite{alazab2021federated}, Alazab et al. explored how FL could enhance authentication, privacy protection, trust management, and attack detection, presenting the critical role FL plays in safeguarding distributed environments. Extending this perspective, recent surveys by Zhang et al.~\cite{zhang2024survey} and Tariq et al.~\cite{tariq2024trustworthy} emphasized the importance of developing trustworthy FL, which incorporates three fundamental principles: ensuring privacy through secure and legally compliant data handling, maintaining security to guarantee confidentiality and accuracy, and promoting fairness by equitably considering client contributions and model inputs.

Building upon this conceptual foundation, Tariq et al.~\cite{tariq2023trustworthy} studied a taxonomy structured around three primary pillars of trustworthy FL: interpretability, fairness, and security and privacy. They suggest that future research should focus on trustless solutions, moving away from reliance on centralized entities to enhance the robustness and resilience of FL systems.

To better understand FL's broader context within deep learning, Almutairi et al.~\cite{almutairi2023federated} compared three prominent training paradigms: centralized training, distributed training, and FL. Their analysis provided a clarified definition of critical FL components, including participant roles, learning processes, aggregation algorithms, partitioning strategies, and data distribution techniques. Moreover, they categorized potential threats to FL systems into two main types: poisoning attacks (covering model and data poisoning) and inference attacks (including reconstruction and membership inference attacks), highlighting the importance of protective strategies for robust and secure FL implementation.

Complementing these perspectives, Mothukuri et al.~\cite{mothukuri2021survey} offered an extensive classification of FL systems, outlining the considerations for building effective FL environments. Their work detailed network topologies, data availability, and partitioning strategies. They further discussed aggregation and optimization algorithms, specifically designed to optimize communication bandwidth, reduce operational costs, and improve aggregation efficiency.

\subsection{About This Survey}
Distinct from traditional FL security research, this study examines novel feature-based threats and their defenses within ResFL. It focuses on poisoning, inference, and reconstruction attacks that exploit benign model features to undermine model integrity and compromise data privacy. To address poisoning threats, the research explores resilient aggregation methods and anomaly detection techniques designed to identify and exclude malicious updates. In addition, it investigates the application of privacy-enhancing technologies, including differential privacy, homomorphic encryption, and secure multi-party computation, to strengthen ResFL within CyberEdge networks.

\section{Joint Training and Agglomerative Deduction}
\label{sec_iii}
This section investigates joint training and agglomerative deduction, which leverages heterogeneous data and hierarchical aggregation with fault tolerance and anomaly detection to improve FL resilience.

\begin{table*}[ht]
\centering
\caption{Heterogeneous Data and Hierarchical Aggregation in ResFL}
\begin{tabular}{|p{3cm}|p{4cm}|p{4cm}|p{4cm}|}
\hline
& \textbf{Representative techniques} & \textbf{Technical specialties} & \textbf{Requirements or limitations} \\
\hline
\textbf{Hierarchical aggregation} & Decentralized peer-to-peer FL~\cite{zhou2023decentralized}, multi-layer aggregation~\cite{you2022reschedule} & Mitigate the impact of data heterogeneity by grouping users for aggregation & Increased communication overhead and complexity in managing multiple layers of aggregation \\
\hline
\textbf{Personalized FL} & Retrogress-resilient FL~\cite{chen2022personalized} & Adapt to user-specific data distributions, reducing performance degradation & May reduce generalization due to excessive personalization \\
\hline
\textbf{Adaptive weighting and rescheduling} & Gradient rescheduling~\cite{you2022reschedule}, adaptive weighting~\cite{zuo2024byzantine}, straggler-resilient FL~\cite{reisizadeh2022straggler} & Improve convergence in non-IID settings and enhance robustness to malicious users & Requiring additional computation and tuning to balance adaptability and stability \\
\hline
\textbf{Federated reinforcement learning} & Vertical federated RL~\cite{mukherjee2024resilient} & Enhance decision-making in cyber-physical systems, such as smart grids & Limited applicability other than reinforcement learning-based tasks \\
\hline
\textbf{CyberEdge FL architectures} & Edge-integrated decentralized FL~\cite{zhou2023decentralized} & Improve privacy and resilience in mobile and distributed systems & Requiring reliable peer-to-peer communication and additional security mechanisms \\
\hline
\end{tabular}
\label{table_fl_heterogeneity}
\end{table*}

\subsection{Heterogeneous Data and Hierarchical Aggregation}
FL inherently operates with heterogeneous data due to the diversity of clients, fluctuating network conditions, and varying application requirements~\cite{yang2022flash,zhou2021two,pang2020realizing}. Addressing this heterogeneity is critical to maintaining the robustness of distributed nodes and ensuring efficient model training~\cite{pei2022tkagfl}. FL can explore hierarchical aggregation as an effective strategy to mitigate the negative impacts of data heterogeneity, thereby preserving model robustness and accuracy.

Data heterogeneity typically manifests in two distinct forms: non-independent and identically distributed (non-IID) data, where the client's local data distributions vary significantly; and system heterogeneity, where participating nodes differ considerably in computational capabilities and communication resources. These factors can adversely affect FL by reducing convergence speed and degrading overall performance~\cite{zheng2022aggregation, chen2022personalized}. Traditional FL aggregation methods, such as FedAvg, often fail to handle heterogeneous data effectively, necessitating the development of more adaptive and robust aggregation strategies~\cite{mukherjee2024resilient}.

Hierarchical aggregation addresses these challenges by organizing clients into subgroups based on factors such as statistical similarity, geographical proximity, or computational capacity. Local models within these clusters are aggregated first, forming an intermediate aggregation layer before global model updates. This additional aggregation step reduces the impact of extreme model divergence caused by non-IID data distributions~\cite{liu2022joint}. 

\begin{figure}[htb]
\centering
\includegraphics[width=\linewidth]{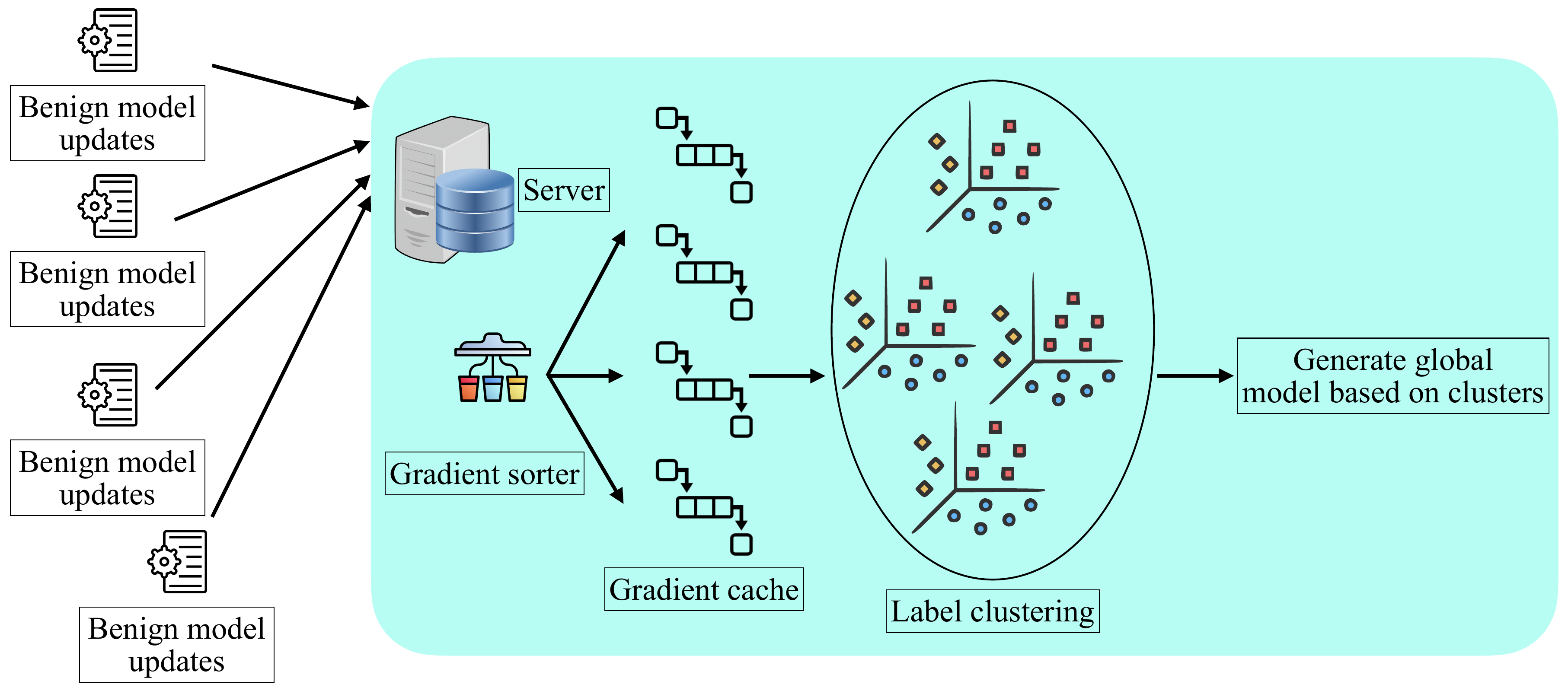}
\caption{Federated gradient scheduling for improving model convergence and stability in heterogeneous environments.}
\label{figure_fedgs}
\end{figure}

Recent studies confirm that multi-tier FL architectures significantly enhance robustness against Byzantine failures and slow-converging clients~\cite{you2022reschedule}. 
For instance, as shown in Fig.~\ref{figure_fedgs}, You et al.~\cite{you2022reschedule} introduced a method of gradient rescheduling that enhances convergence rates and model stability in scenarios with heterogeneous data by arranging the order of gradient updates. Specifically, their approach groups clients based on similarities in label distributions, subsequently reassigning client identities according to these clusters. From these grouped gradients, representative samples can be selected to form an IID gradient batch, providing optimizers with accurate momentum estimates for improved training effectiveness.

Moreover, personalized FL (PFL) techniques address data heterogeneity by customizing local models for individual clients while leveraging global insights. Chen et al.~\cite{chen2022personalized} introduced retrogress-resilient FL methods tailored to handle imbalanced data distributions commonly found in medical applications. These approaches dynamically adapt local model updates based on data disparities, significantly reducing performance degradation associated with heterogeneous data.

Enhanced aggregation frameworks such as gradient rescheduling and adaptive weighting further strengthen FL robustness, specifically against stragglers and adversarial clients. Reisizadeh et al.~\cite{reisizadeh2022straggler} developed a framework that dynamically adjusts client contributions based on reliability metrics, thus balancing statistical accuracy and system heterogeneity to improve overall convergence. Similarly, Zuo et al.~\cite{zuo2024byzantine} presented a Byzantine-resilient FL strategy incorporating adaptive weighting mechanisms that effectively mitigate the influence of malicious clients and enhance system resilience.

CyberEdge networks, which combine edge computing with cyber-physical systems, particularly benefit from hierarchical FL strategies. For instance, Zhou et al.~\cite{zhou2023decentralized} described a decentralized peer-to-peer FL framework for mobile robotic systems, enhancing privacy and resilience through secure, reputation-based virtual clustering. Similarly, Mukherjee et al.~\cite{mukherjee2024resilient} utilized vertical federated reinforcement learning (FedSAC) to optimize energy distribution in smart grids while ensuring robustness against cyber threats, demonstrating its practical applicability through hardware-in-the-loop simulations.

Hierarchical aggregation thus emerges as a promising solution to the challenges posed by data heterogeneity in FL. Future research directions include: dynamic cluster formation to adapt aggregation hierarchies based on real-time data analytics; hybrid aggregation methods combining multi-tier aggregation and reinforcement learning for optimal updates; and enhanced security measures to develop Byzantine-resilient aggregation mechanisms suitable for adversarial environments.
Addressing these challenges will significantly enhance the resilience and adaptability of FL in CyberEdge networks. 

Table~\ref{table_fl_heterogeneity} provides a comprehensive comparison of existing heterogeneous data handling techniques and hierarchical aggregation approaches in ResFL.
In general, FL encounters data heterogeneity stemming from non-IID local distributions and varying node capabilities. This diversity impedes convergence, lowers model accuracy, and demands more robust aggregation methods than conventional approaches like FedAvg. Hierarchical aggregation, which groups clients by factors such as statistical similarity or computing capacity, emerges as a key strategy. Techniques like gradient rescheduling and adaptive weighting further bolster FL’s resilience to Byzantine failures, stragglers, and malicious clients. PFL complements these efforts by tailoring local models while capitalizing on global insights, thereby improving performance under disparate data conditions.
In CyberEdge networks, integrating hierarchical aggregation with novel security measures (e.g., reputation-based clustering, Byzantine resilience, etc.) has proven effective for privacy, reliability, and resource optimization in real-world scenarios like mobile robotics and smart grids. Continued work on adaptive clustering, hybrid aggregation, and advanced security frameworks is anticipated to strengthen FL’s robustness and adaptability in dynamic or adversarial environments.

\subsection{Fault Tolerance and Anomaly Detection}
In FL, it is important to ensure fault tolerance and anomaly detection to maintain the model's reliability in adversarial and resource-constrained environments. This section explores various strategies to enhance FL resilience against Byzantine attacks, communication failures, and malicious data manipulation.

Byzantine failures occur when malicious or compromised clients provide incorrect model updates, potentially degrading overall model performance. Secure aggregation techniques utilize encryption to realize model updates to protect privacy and endure Byzantine failures. The benefit is to prevent bad contributions from affecting the global model. For example, So et al.~\cite{so2020byzantine} presented a Byzantine resilient framework for secure FL, which implements coded computing and cryptographic techniques to prevent adversarial attacks while maintaining efficiency. Their approach, BREA (Byzantine-Resilient Secure Aggregation), integrates stochastic quantization, verifiable outlier detection, and secure model aggregation to ensure both robustness and privacy. Using these techniques, the framework mitigates the impact of malicious updates while preserving data confidentiality. The authors provide theoretical guarantees on convergence and privacy protection, demonstrating that BREA achieves high accuracy even in adversarial settings. Experimental results validate its effectiveness in real-world FL scenarios. Similarly, Xia et al.~\cite{xia2024byzantine} developed an aggregation scheme for privacy protection to improve robustness for Byzantine clients without affecting the confidentiality of the model.

\begin{table*}[ht]
\centering
\caption{Key Techniques for Fault Tolerance and Anomaly Detection in ResFL}
\begin{tabular}{|p{3cm}|p{4cm}|p{4cm}|p{4cm}|}
\hline
& \textbf{Representative techniques} & \textbf{Technical specialties} & \textbf{Requirements or limitations} \\
\hline
\textbf{Secure aggregation} & BREA~\cite{so2020byzantine}, Privacy-preserving aggregation~\cite{xia2024byzantine} & Ensure privacy and resilience to Byzantine failures & Increased computational overhead due to cryptographic techniques \\
\hline
\textbf{Robust gradient aggregation} & Trimmed mean, coordinate-wise median~\cite{tao2023byzantine}, Trust-based weight distribution~\cite{gouissem2023collaborative} & Filters out adversarial updates and improves FL model accuracy & May discard useful gradients along with malicious ones, reducing learning efficiency \\
\hline
\textbf{Anomaly detection} & Gradient anomaly detection~\cite{wei2021gradient}, Reinforced resilient FL (R2Fed)~\cite{zhang2022r} & Detects and mitigates abnormal behaviors caused by adversarial clients or system faults & Requiring additional computational resources for real-time monitoring \\
\hline
\textbf{Behavioral analysis and intrusion detection} & Deep recurrent reinforcement learning for intrusion detection~\cite{kaur2024intrusion}, Differentially private FL~\cite{xiang2023practical} & Identifies adversarial behaviors and prevents privacy leakage & Requiring to balance trade-offs between privacy and model accuracy \\
\hline
\textbf{Resilience schemes} & Collaborative Byzantine-resilient FL~\cite{gouissem2023collaborative}, Low-complexity robust learning~\cite{gouissem2024low} & Enhances FL security with cooperative cross-validation and low-complexity mechanisms & Requiring user cooperation and additional coordination, which may not always be feasible \\
\hline
\end{tabular}
\label{table_fl_fault_tolerance}
\end{table*}

Robust gradient aggregation methods, such as coordinate-wise median and trimmed mean aggregation, filter out extreme values introduced by adversarial clients, reducing their influence on training convergence. Tao et al.~\cite{tao2023byzantine} designed a Byzantine-tolerant FL framework, which combined resilient aggregation rules to mitigate impacts from malicious updates and improved model accuracy in an adversarial environment. Apart from this, the trust-based weight distribution method improves the weight of trust contributions through a trustworthy point system based on the clients' history, so that the robustness can be improved. Gouissem et al.~\cite{gouissem2023collaborative} studied a collaborative Byzantine-resilient FL method in which clients validate each other's updates, increasing resilience and security in federated environments.

\begin{figure}[htb]
\centering
\includegraphics[width=\linewidth]{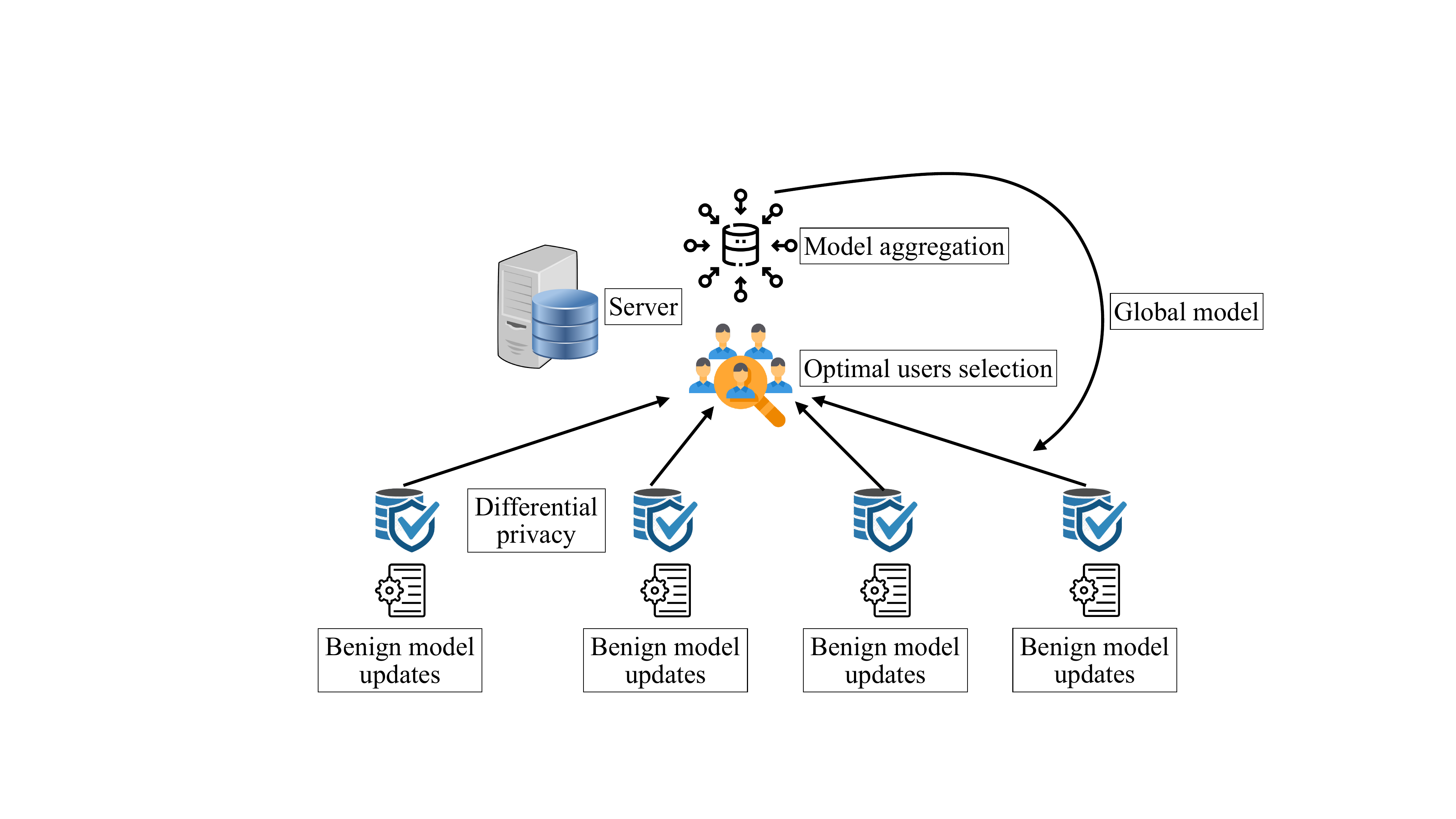}
\caption{Gradient-leak-resistant FL for detecting anomalous gradient patterns and preventing privacy breaches.}
\label{figure_gradientleak}
\end{figure}

Anomaly detection in FL aims to detect and relieve abnormal behavior caused by adversarial clients, connection failures, or hardware malfunctions. The gradient distribution can be analyzed based on abnormal gradient detection to detect and exclude the wrong gradient pattern in the aggregation. As shown in Fig.~\ref{figure_gradientleak}, Wei et al.~\cite{wei2021gradient} developed a gradient-leak-resistant FL method that detects anomalous gradient patterns and prevents privacy breaches while improving overall model robustness. A client-level differential privacy is computed, which adds noise to the shared gradient update by a client at each round of the FL training. Zhang et al.~\cite{zhang2022r} studied a reinforced resilient FL, R2Fed, which is capable of dynamically adjusting the model training strategy given the anomalies detected in the industrial environment, to guarantee the stability of the model's performance.

Behavioral analysis monitors client participation patterns to identify anomalous activity, such as sudden model drift or inconsistent update frequency. Kaur~\cite{kaur2024intrusion} applies deep recurrent reinforcement learning to detect intrusion attempts in industrial Internet of Things networks, demonstrating enhanced anomaly detection capabilities in distributed environments. Furthermore, differential privacy mechanisms introduce noise-based privacy techniques that not only prevent information leakage but also help identify inconsistencies indicative of malicious activity. Xiang et al.~\cite{xiang2023practical} studied a differentially private and Byzantine-resilient FL model that balances security and computational efficiency while maintaining strong privacy guarantees.

In order to reinforce the resilience to FL faults, a framework is presented to integrate redundancy, adaptive learning rate, and reinforcement learning-based resilience mechanisms. R2Fed implements dynamical adjustment of aggregation strategies based on real-time anomaly detection to optimize industrial applications\cite{zhang2022r}. Collaborative Byzantine resilient FL introduces a cooperative learning approach in which clients cross-validate updates of each other before aggregation, improving security and accuracy~\cite{gouissem2023collaborative}. Gouissem et al.~\cite{gouissem2024low} created a low-complexity robust learning mechanism, which reduces computation costs, making Byzantine resilient FL more suitable for resource-limited edge computing.

Advancements in fault tolerance and anomaly detection are important in improving the reliability and security of FL. Future research directions include developing adaptive aggregation frameworks, enabling them to adaptively adjust aggregation rules dynamically based on clients' reliability, implementing deep learning to precisely detect anomalies, and improving encryption protocols to balance the security and efficiency of computing in FL. With these problems solved, FL will be more robust and adaptive in real implementations.

Table~\ref{table_fl_fault_tolerance} compares the representative techniques for fault tolerance and anomaly detection in terms of their specialties, requirements, and limitations.
 
\section{Feature-oriented Threats and Defenses}
\label{sec_iv}
This section studies the feature-oriented threats and defense strategies, which are designed to secure FL models against adversarial manipulations, particularly poisoning attacks, as well as inference and reconstruction attacks. 

\subsection{Feature Extraction with Poisoning Attacks}
\subsubsection{Threat Models}
Model poisoning attacks are a significant threat to FL systems, where adversaries leverage compromised or malicious clients to submit manipulated local updates, intentionally steering the global model away from its correct learning trajectory~\cite{ma2022shieldfl}. The primary objective of these attacks is to mislead the learning outcome, thereby degrading the accuracy, reliability, and trustworthiness of FL-based decision-making.

Several recent studies have expanded the scope and sophistication of model poisoning attacks. For instance, Li et al.~\cite{li2024biasing} introduced the Adversarial Graph Attention Network (AGAT), an advanced adversarial framework specifically designed to launch fairness attacks by strategically manipulating FL training processes. AGAT maximizes the Kullback–Leibler (KL) divergence between user-submitted updates and the global model, utilizing a Graph Autoencoder (GAE) trained via sub-gradient descent to reconstruct correlations among benign model updates. This strategy increases reconstruction loss, ensuring malicious updates remain indistinguishable from genuine contributions, thereby complicating attack detection.

\begin{figure}[h]
\centering
\includegraphics[width=\linewidth]{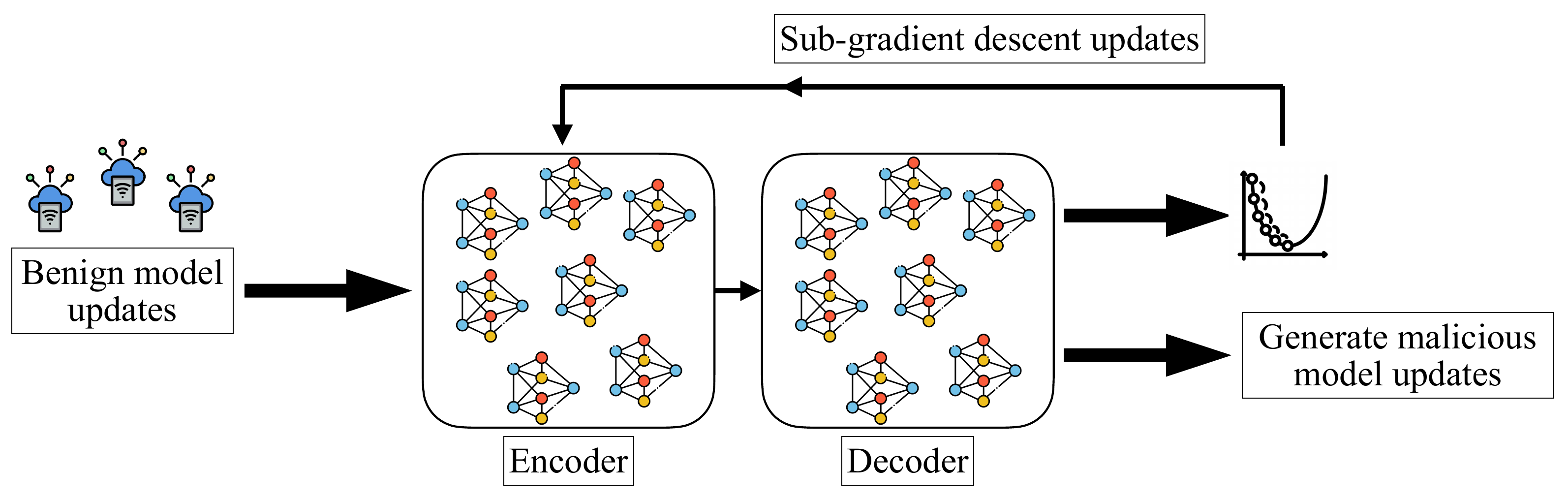}
\caption{Adversarial variational graph autoencoders for crafting malicious models from benign local updates alone, without direct access to training data.}
\label{figure_gae}
\end{figure}

As shown in Fig.~\ref{figure_gae}, a new ``training-data-untethered'' poisoning strategy was proposed by Li et al.~\cite{li2024leverage,li2024data}, which uses adversarial variational graph autoencoders to craft malicious models from benign local updates alone, without direct access to training data. By extracting graph structural correlations and adversarially reconstructing these correlations, the resulting malicious local models become highly effective and particularly challenging to detect, further amplifying threats to FL integrity.

Cao et al.~\cite{cao2022mpaf} developed a distinct approach through a fake-client-based model poisoning attack, where an adversary injects artificially created clients into the FL environment. These fake clients submit deliberately manipulated updates that push the global model towards an adversarially chosen suboptimal baseline, compromising FL system accuracy.

Two poisoning techniques, i.e., label-flipping and model-update poisoning, were systematically examined by Thein et al.~\cite{thein2024personalized} to evaluate their detrimental impact on FL-based intrusion detection systems. The study pointed out a critical weakness: As the heterogeneity of user data increases, robust aggregation methods fail to effectively mitigate poisoned contributions, causing significant performance degradation. Supporting this finding, Abou et al.~\cite{abou2023mitfed} noted that data heterogeneity further exacerbates challenges to global model convergence, making FL systems even more vulnerable to poisoning attacks.

Expanding on traditional poisoning methods, Yang et al.~\cite{yang2023model} introduced a model shuffle poisoning attack that involves strategically shuffling and scaling parameters within malicious models. Unlike conventional approaches, this method preserves benign appearances and test accuracy, subtly disrupting global model convergence. As a result, it can slow down learning or lead to divergence, complicating its identification and mitigation.

Focused specifically on FL in autonomous vehicles, Wang et al.~\cite{wang2023bandit} designed a dynamic data poisoning framework leveraging a bandit-based approach. Their black-box attack adaptively selects vulnerable regions within the steering angle regression task, increasing effectiveness across FL training rounds while evading detection mechanisms.

In addition, backdoor poisoning remains an ongoing concern, as presented by Lyu et al.~\cite{lyu2023poisoning}. Their approach enables malicious actors to insert covert triggers into FL models by coordinating model updates from multiple compromised clients. Specifically designed to bypass common defense strategies, this backdoor attack presents a persistent and stealthy threat capable of severely undermining FL system security. 

\subsubsection{Defense Strategies}
Recent research has increasingly focused on developing defense mechanisms to counter sophisticated model poisoning attacks in FL. Zhang et al. introduced FedCAMAE~\cite{zheng2024exploring}, a new defense approach that leverages visual explanation techniques to enhance detection capabilities beyond conventional Euclidean distance-based or machine learning-based methods. Specifically, FedCAMAE integrates Layer Class Activation Mapping (LayerCAM) with an autoencoder to produce detailed heat maps for each local model update submitted to the central server. These heat maps serve as fine-grained visual representations, which the autoencoder further refines, highlighting hidden features and improving distinguishability between benign and malicious updates.

\begin{table*}[htb]
\centering
\caption{Representative Techniques of Feature-oriented Threats and Defenses Strategies}
\begin{tabular}{|p{3cm}|p{4cm}|p{4cm}|p{4cm}|}
\hline
 & \textbf{Representative techniques} & \textbf{Technical specialties} & \textbf{Requirements or limitations} \\
\hline
\textbf{Graph-based attacks} & AGAT~\cite{li2024biasing}, Variational Graph Autoencoders~\cite{li2024leverage, li2024data} & Effective in stealthily compromising FL by exploiting graph structural correlations & Hard to defend due to the lack of direct data access and complexity in detecting graph-based manipulations \\
\hline
\textbf{Fake client model poisoning} & Fake-client attack~\cite{cao2022mpaf}, Label-flipping attack~\cite{thein2024personalized}, Model shuffle poisoning~\cite{yang2023model} & Can significantly degrade FL performance by injecting manipulated clients & Detecting fake clients is challenging, especially in non-IID settings \\
\hline
\textbf{Adaptive backdoor} & Bandit-based poisoning~\cite{wang2023bandit}, Backdoor poisoning~\cite{lyu2023poisoning} & Adaptive techniques can improve attack efficacy while evading detection mechanisms & Requiring continuous learning to maintain effectiveness and can be computationally expensive \\
\hline
\textbf{Visual and similarity-based defenses} & FedCAMAE~\cite{zheng2024exploring}, GradCAM-AE~\cite{zheng2024detecting} & Improves model security through visual-based anomaly detection & May suffer from increased computational overhead and reliance on feature interpretability \\
\hline
\textbf{Clustering and model filtering defenses} & Representational similarity~\cite{chen2024exploring}, Top-$k$ sparsification~\cite{panda2022sparsefed}, LOMAR~\cite{li2021lomar}, FL ensemble~\cite{cao2022flcert}, RobustFL~\cite{zhang2021robustfl} & Effective in detecting adversarial models through clustering and statistical techniques & Limited effectiveness in highly dynamic adversarial environments with adaptive attacks \\
\hline
\end{tabular}
\label{table_attack_defense}
\end{table*}

\begin{figure}[h]
\centering
\includegraphics[width=\linewidth]{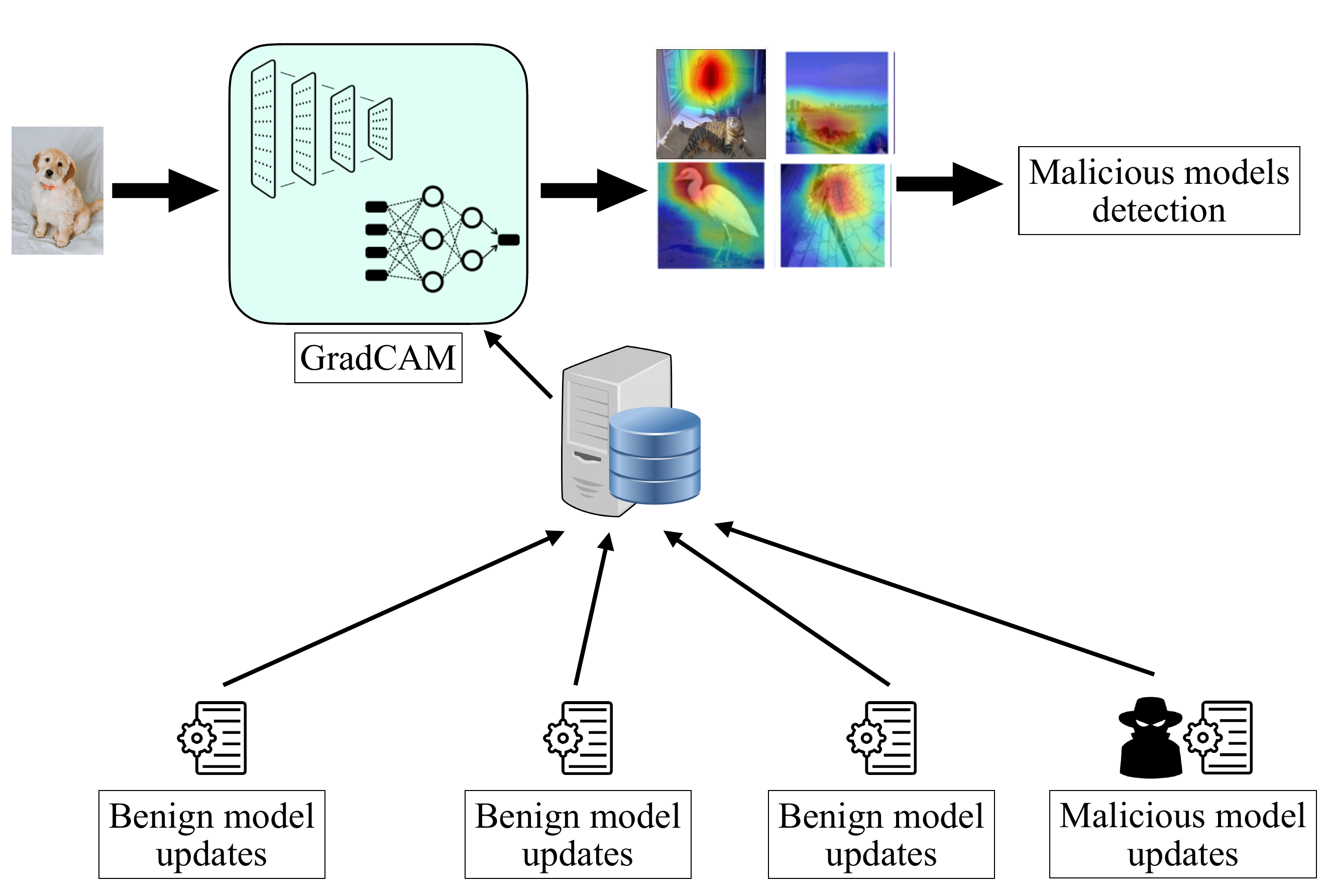}
\caption{Detecting poisoning attacks on FL using GradCAM~\cite{zheng2024detecting}.}
\label{figure_gradcam}
\end{figure}

Extending visual explanation-based defenses, another promising approach was proposed by Zheng et al.~\cite{zheng2024detecting}. As shown in Fig.~\ref{figure_gradcam}, their framework combines Gradient-weighted Class Activation Mapping (GradCAM) and autoencoders, effectively improving detection accuracy against model poisoning attacks. By analyzing anomalous heat maps generated through GradCAM, this method can strengthen FL security, enhancing the ability to identify and isolate malicious updates.

Many existing defenses often assume independent and identically distributed (IID) data environments, resulting in reduced performance when confronted with non-IID data. To address these limitations, Chen et al.~\cite{chen2024exploring} developed a defense mechanism specifically designed for non-IID scenarios. Their method employs representational similarity analysis to systematically evaluate the alignment between global and local models. By constructing a representational consistency set and applying clustering algorithms such as $k$-means, the framework effectively identifies and isolates adversarial entities, improving defense robustness in heterogeneous data settings.

Complementing these visual and clustering-based defenses, Panda et al.~\cite{panda2022sparsefed} introduced SparseFed, a technique that mitigates model poisoning attacks through gradient clipping and top-$k$ sparsification. During each training round, only the top-$k$ gradients with the highest magnitude are aggregated and used to update the global model. Since attackers often manipulate gradients in directions divergent from benign updates, their malicious contributions are inherently minimized or excluded.

A two-phase approach for detecting malicious updates was further extended in~\cite{li2021lomar}. In the first phase, kernel density estimation evaluates the relative distribution of local model updates, identifying anomalous patterns. In the second phase, a statistical detection threshold differentiates malicious from benign updates. This structured analysis enhances the precision of identifying compromised users, significantly strengthening FL resilience.

Cao et al.~\cite{cao2022flcert} expanded the defense landscape by designing an ensemble-based FL framework. Their strategy partitions users into multiple groups, training a separate global model for each group independently. A majority voting mechanism then aggregates predictions from these models, substantially enhancing FL robustness and minimizing the impact of adversarial manipulations.

Focusing specifically on FL applications within IoT environments, Zhang et al.~\cite{zhang2021robustfl} introduced a logits-based predictive model deployed at the server level. This model helps identify and trace incoming logits, effectively pinpointing potentially malicious sources. Concurrently, the federated model undergoes adversarial training, proactively counteracting attacker manipulations and substantially complicating stealthy poisoning attacks.

Table~\ref{table_attack_defense} describes the critical representative techniques of feature-oriented threats and defense strategies.

\subsection{Feature-based Inference and Reconstruction Attacks}

Feature-based inference and reconstruction attacks pose significant threats to FL, as adversaries can exploit shared model updates to infer sensitive information.

\subsubsection{Threat Models}
One significant threat to FL systems is the Generative Adversarial Network (GAN)-based reconstruction attack, which leverages adversarial learning to reconstruct private training data from model updates. Jere et al.~\cite{jere2020taxonomy} categorized various FL attacks, emphasizing that reconstruction and model inversion attacks commonly exploit gradient leakage, enabling adversaries to infer sensitive information directly from parameter updates. Ha and Dang~\cite{ha2022inference} specifically investigated GAN-driven inference attacks, demonstrating that a well-trained GAN could generate highly accurate approximations of the original training data, raising serious privacy concerns in federated environments.

\begin{figure}[t]
\centering
\includegraphics[width=\linewidth]{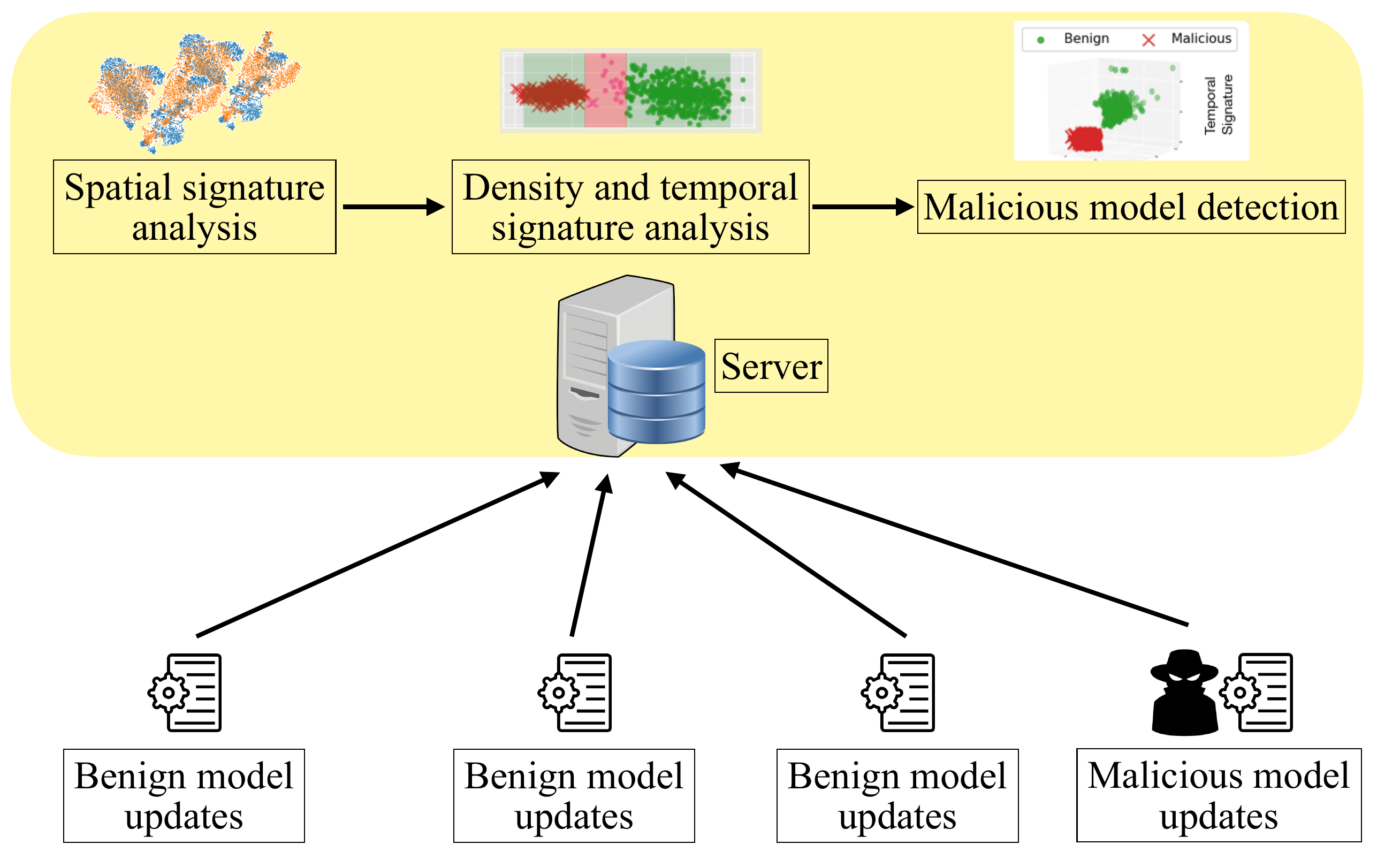}
\caption{Stdlens, a model hijacking-resilient FL for object detection~\cite{chow2023stdlens}.}
\label{figure_stdlends}
\end{figure}

Expanding upon these concerns, Chow et al.~\cite{chow2023stdlens} introduced Stdlens, as shown in Fig.~\ref{figure_stdlends}, a resilient FL framework explicitly designed to mitigate model hijacking and gradient-based reconstruction attacks. Their approach developed a three-tier analysis, including spatial signatures, density, and temporal signatures, and model detection analysis, to reduce vulnerabilities, making it more challenging for adversaries to exploit model updates to recover private information.

Further studies examined various dimensions of inference attacks in~\cite{hu2021source} analyzed source inference attacks, illustrating that adversaries could identify the origins of specific data samples by analyzing model updates without direct access to raw data. Extending this concept, Hu et al.~\cite{hu2023source} later demonstrated that source inference attacks pose deeper privacy risks, surpassing traditional membership inference attacks in severity.

In another work, Luo et al.~\cite{luo2021feature} examined feature inference attacks within vertical FL frameworks, where malicious entities can infer sensitive attributes from encrypted model predictions, highlighting vulnerabilities even when data is securely partitioned. Gao et al.~\cite{gao2021secure} further noted that standard secure aggregation techniques alone are insufficient in protecting FL systems from category inference attacks, indicating potential weaknesses in existing security protocols.

\begin{table*}[htb]
\centering
\caption{Key Techniques of Inference and Reconstruction Attacks and Defending Models in ResFL}
\begin{tabular}{|p{3cm}|p{4cm}|p{4cm}|p{4cm}|}
\hline
& \textbf{Representative techniques} & \textbf{Technical specialties} & \textbf{Requirements or limitations} \\
\hline
\textbf{Secure aggregation and encryption} & Private aggregation scheme~\cite{zhao2022practical}, ARM TrustZone~\cite{messaoud2022shielding} & Enhancing privacy protection by securing model updates & Increased computational overhead may impact ResFL efficiency \\
\hline
\textbf{Adversarial perturbation} & Defensive neural networks~\cite{lee2021defensive}, users-level input perturbation~\cite{yang2023fortifying}, adversarial examples~\cite{xie2021defending} & Obfuscate gradients to prevent inference attacks & May degrade model accuracy due to added noise \\
\hline
\textbf{Differential privacy-based defenses} & ResFL models with differential privacy~\cite{xu2024robust}, user-level differential privacy~\cite{feng2022user} & Limit adversaries' ability to infer sensitive data while preserving utility & Privacy-utility trade-off may affect model performance \\
\hline
\textbf{Gradient perturbation and secure learning} & FLSG~\cite{fan2023flsg}, Gradient perturbation techniques~\cite{feng2022user} & Reduce adversaries' ability to extract private features from gradients & Requiring careful tuning to balance security and convergence \\
\hline
\textbf{Hardware-Based security mechanisms} & ARM TrustZone-based protection~\cite{messaoud2022shielding} & Strengthens privacy with secure enclaves and hardware isolation & Implementation complexity and hardware dependency \\
\hline
\end{tabular}
\label{table_fl_inference_defenses}
\end{table*}

\begin{figure}[h]
\centering
\includegraphics[width=\linewidth]{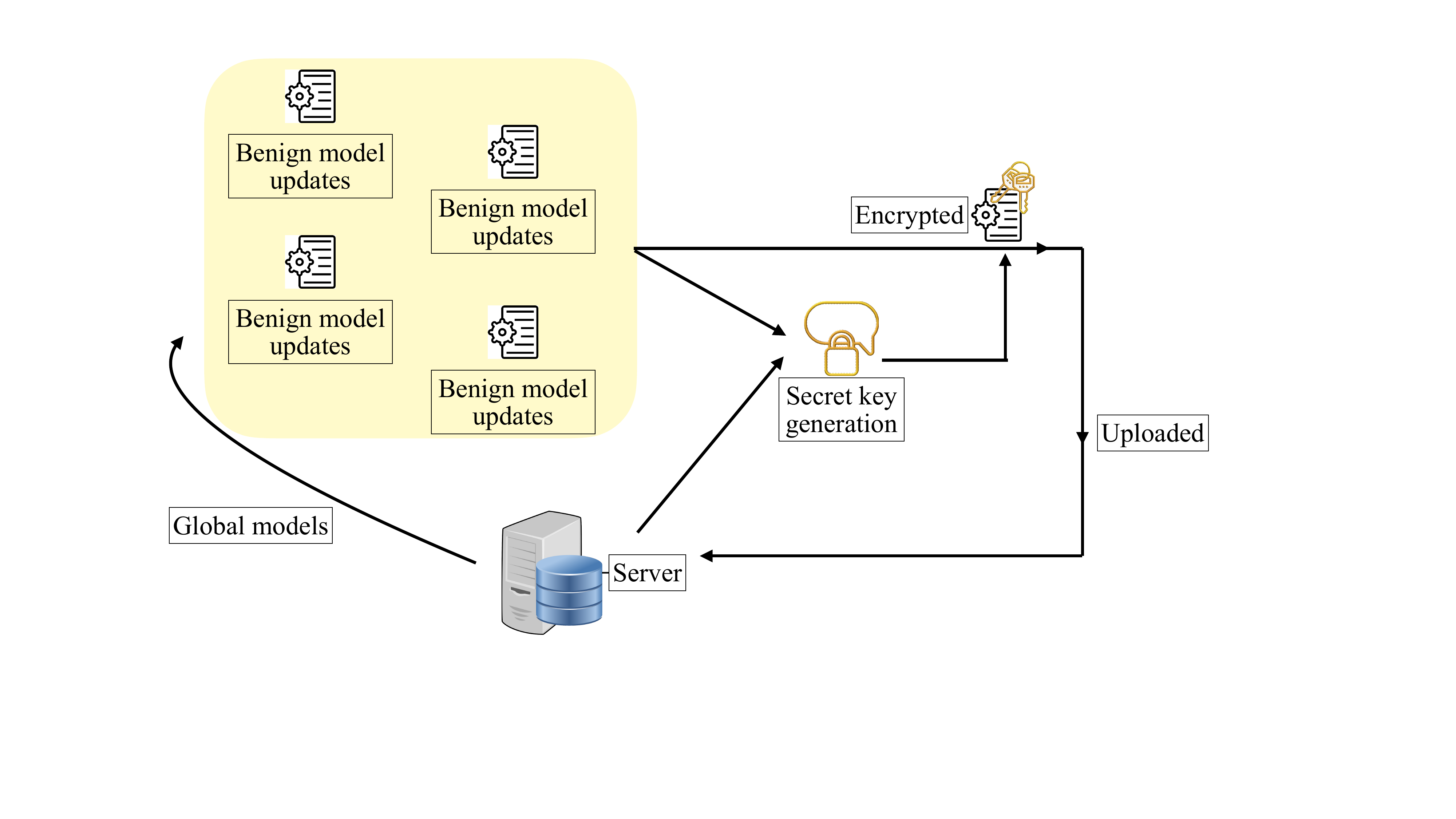}
\caption{A private aggregation scheme in FL against inference attacks~\cite{zhao2022practical}.}
\label{figure_privateagg}
\end{figure}

Additional vulnerabilities were identified in~\cite{fu2022label}, where label inference attacks were studied in vertical FL scenarios, revealing how adversaries could recover sensitive labels from federated models. Wang et al.~\cite{wang2022poisoning} showed that adversarial data manipulation, through poisoning-assisted property inference attacks, could facilitate privacy breaches, demonstrating how poisoning attacks directly enable inference risks. Moreover, Yang et al.~\cite{yang2023practical} presented a practical feature inference attack targeting real-world FL deployments in artificial intelligence of things environments, underscoring significant operational vulnerabilities.

Given the evolving sophistication of GAN-based and other reconstruction attacks, it is critical to develop robust countermeasures. Strengthening differential privacy techniques, improving secure aggregation protocols, and incorporating adversarial training are key areas for enhancing FL's resistance to inference attacks, thus safeguarding data privacy in federated systems.

\subsubsection{Defense Strategies}
To counter the rising threat of feature-based inference and reconstruction attacks, researchers have developed various defensive strategies aimed at preserving privacy and enhancing the robustness of FL systems. Zhao et al.~\cite{zhao2022practical} introduced a private aggregation scheme that can strengthen FL systems against inference attacks. As shown in Fig.~\ref{figure_privateagg}, their approach leverages advanced encryption techniques, securing model updates effectively without compromising computational efficiency. Complementing this, Lee et al.~\cite{lee2021defensive} explored defensive neural networks employing adversarial perturbation techniques designed to obfuscate gradients, thus making data reconstruction substantially more difficult for adversaries.

Further advancing privacy protections, Xu et al.~\cite{xu2024robust} integrated differential privacy mechanisms directly into FL models, reducing vulnerability to client-side data inference attacks. Their approach demonstrates how robust modeling practices can limit adversarial inference capabilities. In particular, Fan et al.~\cite{fan2023flsg} introduced FLSG, a defense specifically tailored for vertical FL scenarios, utilizing gradient perturbation to impede adversaries from extracting sensitive feature information during model training.

Addressing user-level vulnerabilities, Feng et al.~\cite{feng2022user} presented a differential privacy method tailored explicitly for speech emotion recognition models in FL environments. Their technique prevents adversaries from reliably inferring personal attributes from model data, demonstrating effectiveness in practical user-level privacy scenarios. Moreover, hardware-based security solutions were explored by Messaoud et al.~\cite{messaoud2022shielding}, who demonstrated the feasibility of ARM TrustZone technology for safeguarding FL systems from inference attacks through trusted execution environments.

Expanding these defensive methodologies, Yang et al.~\cite{yang2023fortifying} developed client-level input perturbation techniques specifically designed to resist membership inference attacks. Concurrently, Xie et al.~\cite{xie2021defending} demonstrated the effectiveness of adversarial examples as a protective measure, strategically obfuscating sensitive data to counter inference threats.

In addition, Table~\ref{table_fl_inference_defenses} compares the pros and cons of the key techniques of inference and reconstruction attacks and defending models in ResFL.

\section{Opportunities of ResFL}
\label{sec_v}
\begin{figure*}[h]
\centering
\begin{tabular}{cc}
\includegraphics[height=0.34\textwidth]{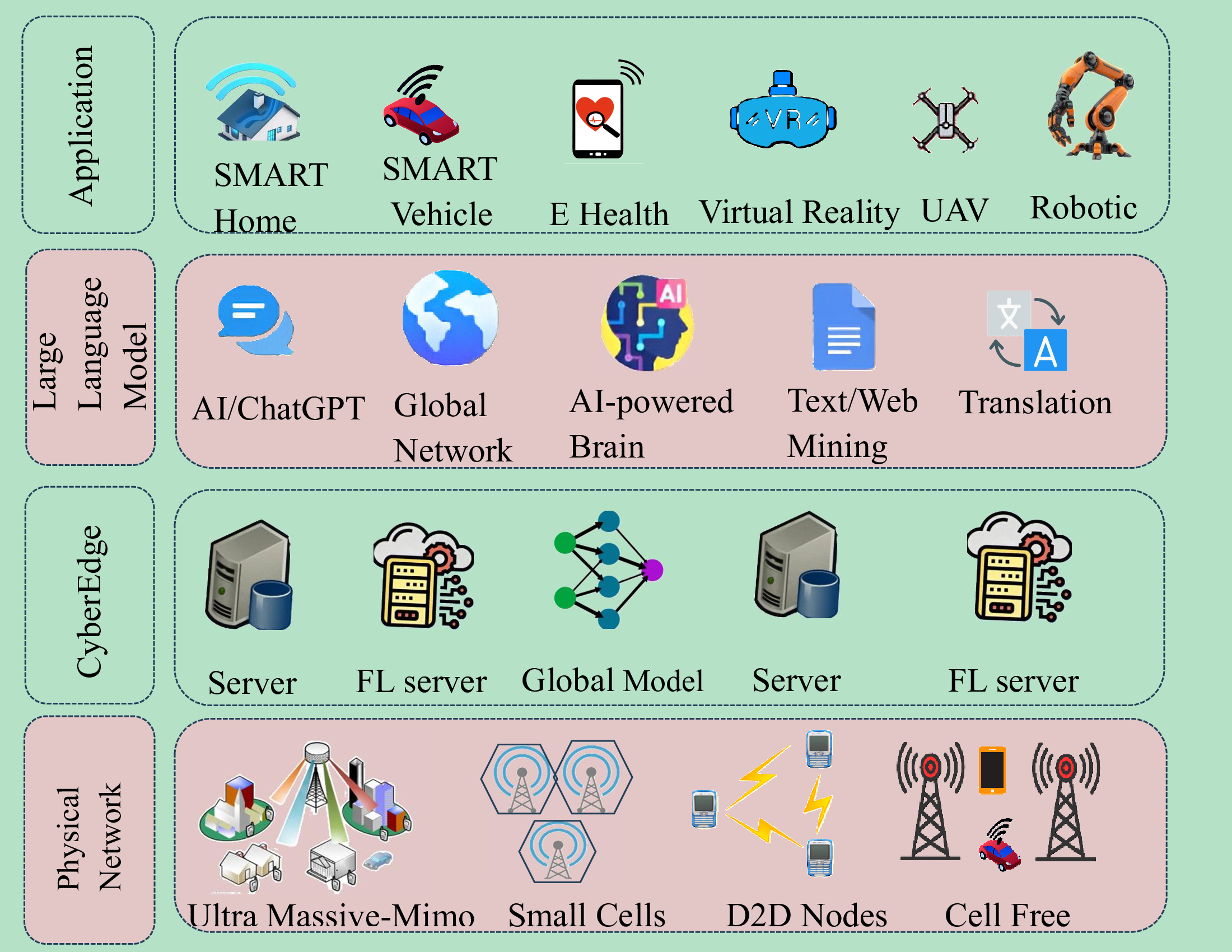} & \includegraphics[height=0.34\textwidth]{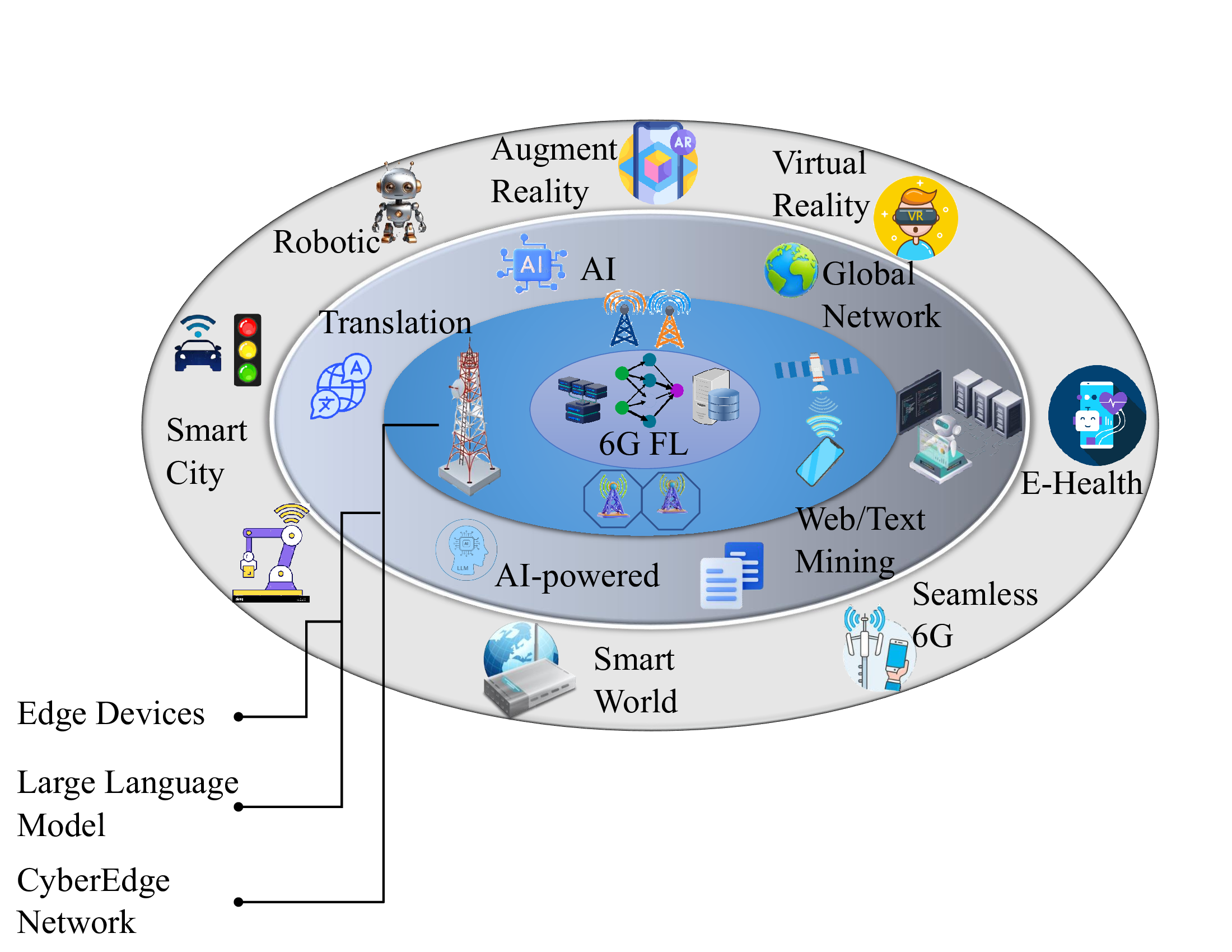}
\end{tabular}
\caption{Research opportunities for developing future ResFL in CyberEdge networks.}
\label{figure_opportunities}
\end{figure*}

In this section, we explore research opportunities for developing future ResFL, including achieving an optimal balance between communication efficiency and model training performance, as well as designing scalable hierarchical aggregation, as illustrated in Fig.~\ref{figure_opportunities}. Overcoming these challenges is crucial for establishing a connected and trustworthy environment that ensures high resilience for users in CyberEdge networks.

\subsection{6G-Assisted ResFL}
One of the primary challenges in 6G-assisted ResFL is achieving an optimal trade-off between communication efficiency and model training performance. While 6G offers ultra-low latency and high-speed data transmission to CyberEdge networks, FL models, especially large-scale deep learning models, still require significant communication resources for frequent parameter exchanges between edge devices and servers~\cite{jiang2025enhancing,zhang2025security,li2017wireless}. The challenge intensifies when considering device mobility, fluctuating network conditions, and limited energy budgets. 

Future work needs to develop adaptive communication strategies such as event-triggered updates, sparsification, quantization, and hierarchical aggregation to reduce transmission overhead while maintaining model accuracy and robustness. Moreover, integrating semantic communication in FL, where only the most informative features are transmitted rather than raw updates, could further enhance efficiency in 6G environments.

Another challenge in 6G-assisted ResFL is the new management of heterogeneous data distributions and the design of scalable hierarchical aggregations to improve learning efficiency as well as resilience. In CyberEdge networks, edge devices generate diverse data types with varying quality, availability, and statistical distributions, making it difficult to achieve global model generalization while maintaining local adaptability. 6G's AI-native infrastructure can enable intelligent data clustering, adaptive aggregation, and cross-layer coordination~\cite{chiarani2025xai,hasan2024federated,raja2024efficient}, but optimally selecting aggregation points and balancing local versus global updates remain open problems. 

As a next-step direction, it is critical to explore dynamic hierarchical aggregation mechanisms that can be adjusted based on network conditions, device reliability, and data distribution patterns. In addition, integrating federated meta-learning and transfer learning can help ResFL models quickly adapt to new environments and unseen data distributions while reducing computational and communication burdens in large-scale, hierarchical 6G-supported CyberEdge networks.

\subsection{Joint Training of LLMs and ResFL}
LLMs can enable privacy-preserving and decentralized training of ResFL across edge devices while maintaining robustness~\cite{kuang2024federatedscope,wu2024fedbiot,wang2025design}. However, integrating LLMs with ResFL in CyberEdge networks can introduce considerable security risks, particularly in defending against adversarial threats, such as model poisoning, backdoor attacks, and inference attacks. Since LLMs require extensive training on diverse datasets, attackers can exploit their federated nature by injecting malicious updates, subtly altering model behavior, or embedding hidden backdoors that trigger harmful outputs under specific conditions. Unlike conventional FL models, LLMs are more susceptible to memorization and prompt-based vulnerabilities, increasing the risk of data leakage even in privacy-preserving settings. 

Future research is required to focus on robust defense mechanisms, including adversarial training, anomaly detection, and secure aggregation techniques, to identify, isolate, and mitigate adversarial influences in ResFL deployments. 

On the other hand, maintaining trustworthiness in the joint training of LLMs and ResFL is challenging due to the heterogeneous and decentralized nature of edge devices, each contributing to updates that may vary in quality, reliability, or intent~\cite{cheng2024towards,friha2024llm,han2024fedsecurity}. In particular, LLMs require complex semantic understanding, making them prone to biased learning, inconsistent generalization, and unreliable knowledge aggregation in diverse CyberEdge networks~\cite{liu2025differentially,wu2024fedbiot,hu2024federated}. Furthermore, verifying the integrity of local updates and preventing misinformation propagation become critical issues, especially when LLMs are applied in sensitive applications, such as autonomous systems, healthcare, and finance. 

More efforts are needed to develop trust-aware ResFL frameworks, integrating advanced techniques, including blockchain-based verification, reputation-based client scoring, and incentive-driven participation mechanisms, to ensure fair and reliable model contributions. In addition, self-assessment strategies within LLMs can be explored to evaluate their own responses for potential biases or hallucinations, enhancing overall trust in ResFL-based decision-making systems.

\subsection{Cross-Domain and Cross-Silo ResFL}
Collaborative cross-domain and cross-silo ResFL presents a critical opportunity for enhancing resilience in future CyberEdge networks, especially across sectors, such as healthcare, autonomous vehicles, finance, and smart cities~\cite{wang2025federated,tian2024privacy,wang2024fedccrl,zhang2024distributed,ali2024federated}. In these contexts, data is inherently fragmented across various entities or domains, such as hospitals, automotive manufacturers, or municipal infrastructures, where each maintains distinct data characteristics and strict privacy constraints. Enabling these organizations to collaboratively train ResFL models without sharing raw data can unlock powerful intelligence while preserving data sovereignty. However, such collaboration is non-trivial, as it has to overcome challenges related to data heterogeneity, system interoperability, trust, and compliance with domain-specific regulations.

A major research direction is the development of interoperable learning frameworks that allow cross-domain ResFL systems with differing data modalities, model architectures, and system capabilities to participate effectively in joint training. This involves designing flexible ResFL protocols that support asynchronous updates, heterogeneous model fusion, and hybrid aggregation strategies adaptable to varying data formats and tasks (e.g., combining image-based diagnostics from healthcare with numerical sensor data from vehicular networks)~\cite{xie2024efficiency,zhao2022federated,liu2022deep,li2019energy}. Such frameworks should also account for resource diversity, enabling both high-end servers and lightweight edge devices to contribute proportionally without compromising the global model. Building standardized APIs and modular ResFL interfaces across platforms will be essential for seamless integration and deployment at scale.

Another key direction lies in robust domain adaptation techniques tailored for ResFL settings. Since data distributions often differ significantly across domains (non-IID data)~\cite{borazjani2025redefining,li2021fedrs}, global models trained via conventional FL may suffer from poor generalization. Future research can investigate domain-invariant feature extraction, personalized ResFL, and meta-learning methods to allow global models to learn from cross-domain knowledge while adapting to local nuances. Moreover, hierarchical ResFL and clustered ResFL approaches can be leveraged to group similar domains before federating at a higher level, improving convergence speed and performance while preserving domain-specific insights. These solutions should also be resilient to domain shifts and adversarial conditions, ensuring stable performance in real-world dynamic environments.

In addition, the design of privacy-preserving and regulation-compliant protocols is essential for trusted cross-silo ResFL. Each domain or organization may operate under different legal and ethical standards (e.g., GDPR in Europe, or HIPAA in the US), requiring tailored privacy guarantees~\cite{wang2024social,zhang2024ppfed,xu2024adaptive}. Advanced techniques, such as differential privacy, secure multi-party computation, federated analytics, and trusted execution environments, will play an important role in enabling secure model training without compromising sensitive data. Moreover, auditable federated mechanisms using blockchain or distributed ledgers could ensure accountability and trust among participating silos. By addressing these challenges, cross-domain and cross-silo ResFL can empower CyberEdge networks with high resilience, scalability, and security across diverse and decentralized ecosystems.

\section{Conclusions}
\label{sec_vi}
This survey focused on ResFL in CyberEdge networks, which is a rapidly evolving field that demands novel approaches to enhance security, efficiency, and adaptability. We explored feature-oriented threats, such as poisoning, inference, and reconstruction attacks, which remain critical, requiring continuous advancements in anomaly detection and resilient aggregation techniques. We investigated adaptive hierarchical learning and fault tolerance mechanisms that play a crucial role in mitigating the challenges posed by non-IID data and unreliable devices, ensuring stable convergence and efficient communication. For future opportunities, the incorporation of 6G and LLMs offers significant potential for improving decentralized and privacy-preserving learning, leveraging ultra-low latency, massive connectivity, and AI-driven optimization. ResFL can also pave the way for robust cross-domain and cross-silo edge intelligence, such as autonomous systems, healthcare, and smart cities, where data privacy and resilience are paramount. As future research unfolds, interdisciplinary collaboration among security, networking, and AI communities will be a key to realizing the full potential of ResFL and driving its real-world deployment.



\ifCLASSOPTIONcaptionsoff
  \newpage
\fi

\bibliographystyle{IEEEtran}
\bibliography{bibCE}

\end{document}